\documentclass[useAMS,usenatbib]{mn2e}

\usepackage{graphicx}

\def\Lsun{L$_\odot$}
\def\Msun{M$_\odot$}
\def\Mbh{$M_{\rm BH}$}

\def\Oiii{[O\,{\sc iii}]}
\def\Oii{[O\,{\sc ii}]}

\def\kms{km\,s$^{-1}$}

\def\lsim{\mathrel{\rlap{\lower 3pt \hbox{$\sim$}} \raise 2.0pt \hbox{$<$}}}
\def\gsim{\mathrel{\rlap{\lower 3pt \hbox{$\sim$}} \raise 2.0pt \hbox{$>$}}}
\def\apj{ApJ}
\def\apjl{ApJL}
\def\mnras{MNRAS}

\title[New insights on J0927+2943]{New insights on the recoiling/binary black hole candidate J0927+2943 via molecular gas observations}
\author[Decarli et al.]{
R. Decarli$^{1}$\thanks{E-mail: decarli@mpia.de}, M. Dotti$^{2,3}$, C. Mazzucchelli$^{1,2}$, C. Montuori$^{4}$, M. Volonteri$^{5}$\\
$^{1}$Max-Planck Institut f\"{u}r Astronomie, K\"{o}nigstuhl 17, D--69117, Heidelberg, Germany\\
$^{2}$Dipartimento di Fisica G. Occhialini, Universit\`{a} degli Studi di Milano--Bicocca, Piazza della Scienza 3, I--20126 Milano, Italy\\
$^{3}$INFN, Sezione Milano-Bicocca, Piazza della Scienza 3, I--20126 Milano, Italy.\\
$^{4}$Dipartimento di Fisica e Matematica, Universit\`{a} dell'Insubria, via Valleggio 11, I--22100 Como, Italy\\
$^{5}$Institut d'Astrophysique de Paris, 98 bis Bd Arago, Paris, F--75014, France}
\begin{document}


\pagerange{\pageref{firstpage}--\pageref{lastpage}} 

\maketitle

\label{firstpage}

\begin{abstract}
The peculiar QSO J0927+2943 shows multiple sets of emission lines in its optical spectrum. This signature has been interpreted as the relative motion between a black hole, either recoiling or bound in a binary system, and its host galaxy, or as a superposition of two galaxies along the line of sight. In order to test these scenarios, we have collected 2mm CO(2--1) observations using the IRAM Plateau de Bure Interferometer, and optical images and spectroscopy at the Calar Alto observatory. Together with archival {\em HST} images, these data provide unique insights on the nature of this system. The recoiling/binary black hole scenarios are ruled out by the clear detection of a galactic--scale molecular gas reservoir at the same redshift of the QSO broad lines. The observations presented here also disfavour the superposition model, although with less constraints. Thus, the origin of the second, bright set of narrow emission lines in J0927+2943 is still unknown.
\end{abstract}


\section{Introduction}

\begin{figure*}
\includegraphics[width=0.99\textwidth]{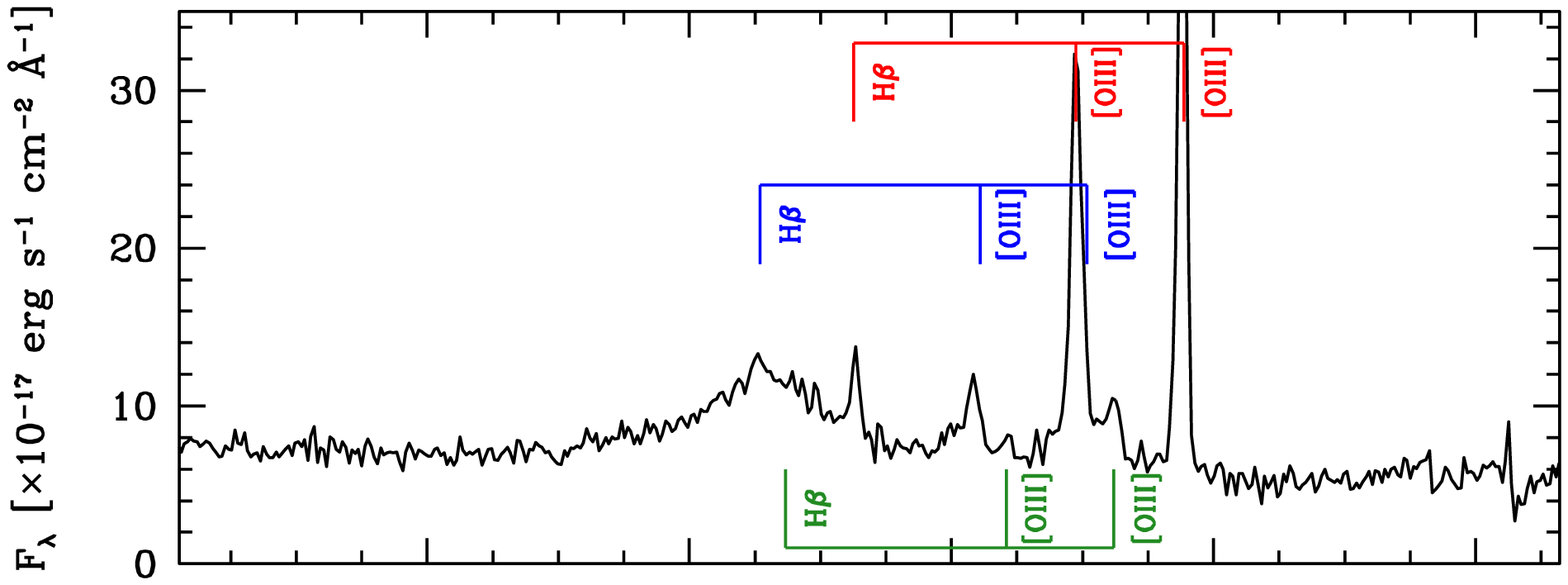}\\
\includegraphics[angle=-90, width=0.99\textwidth]{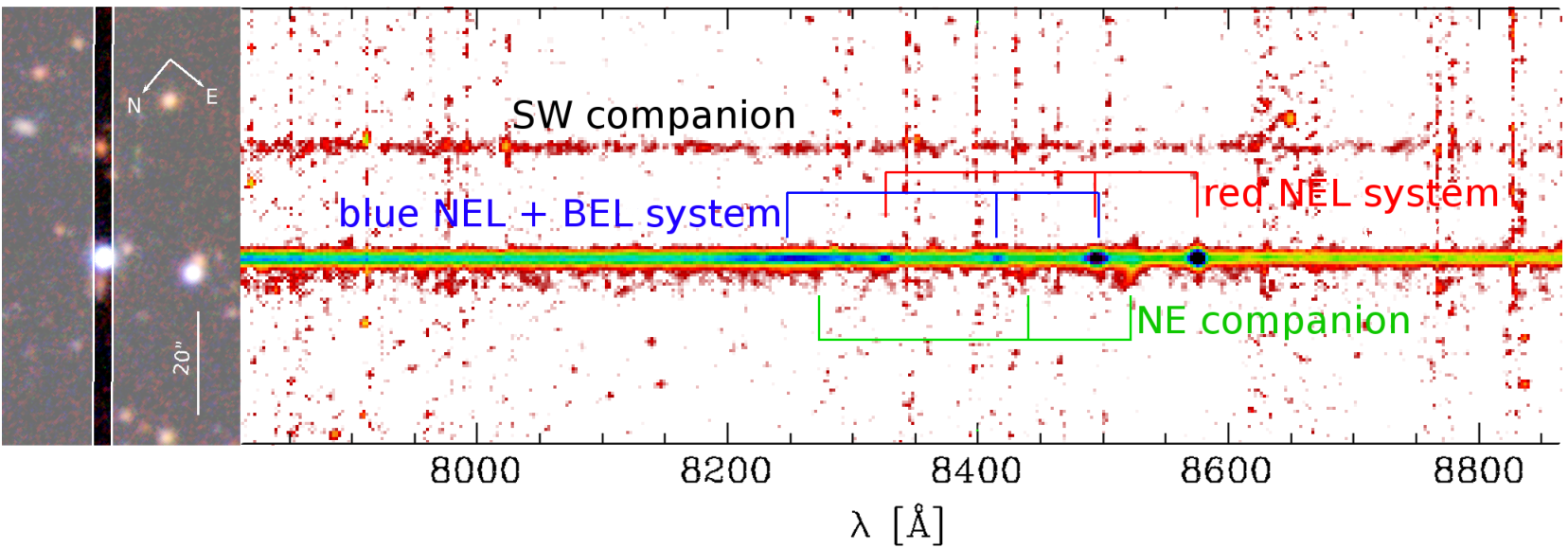}
 \caption{A detail of the optical spectrum of J0927+2943 taken with the Calar Alto 3.5m telescope. {\em Top panel ---} The 1-D spectrum extracted along the QSO continuum. The relevant redshift systems and the associated emission lines are labeled. {\em Bottom left panel ---} Postage stamp of J0927+2943, showing the orientation of the slit used in our spectroscopic observations. The RGB channels are our BUSCA $I$-, $R$-, and $B$-band images, respectively. {\em Bottom right panel ---} The 2-D, sky--subtracted spectrum. The slit orientation is such that NE in at the bottom, to SW is at the top. The bright set of narrow lines (`red NEL') is spatially consistent with the continuum of the QSO down to sub-pixel scales, and is spatially unresolved at seeing scales. The \Oiii{} emission of the NE companion on the other hand is clearly extended all the way towards the QSO (and possibly even beyond). The \Oiii{} redshift is consistent with the one of the CO(2--
 1) emission.}\label{fig_caha_spc}
\end{figure*}

The spectral properties of the quasar SDSS J092712.65+294344.0 (hereafter: J0927+2943) have attracted much attention since its first recognition by \cite{Komossa08}. Its spectral peculiarity is due to the presence of emission lines at different redshifts: a set of broad and (faint) narrow emission lines (BELs and NELs respectively) is observed at $z=0.697$ (hereafter, the `blue' system), together with a set of brighter NELs at $z=0.7128$ (the `red' system, corresponding to a velocity difference $\Delta v=2650$ \kms{}; see Fig.~\ref{fig_caha_spc}).

\cite{Komossa08} proposed J0927+2943 as a first example of recoiling AGN candidate, i.e., the result of the coalescence of a massive black hole (BH) binary (e.g. \citealt{Campanelli07}, \citealt{Lousto11}).  In this scenario, the BH remnant is kicked out of the host nucleus at a velocity of up to $\sim 5000$ km s$^{-1}$. Such a high speed allows the BH to drag along only matter initially very close to it, namely the accretion disc (if present) and the gas responsible for the BELs (\citealt{Loeb07}, \citealt{Komossa08}). Gas orbiting at larger distances, such as the molecular torus and gas emitting the NELs, is not gravitationally bound to the recoiling BH and cannot follow the BH motion (e.g. \citealt{Lusso14}). Hence the NELs can be used as tracers of the host galaxy recessional velocity. In this scenario, the redshift difference between the blue and red sets of lines is caused by relative velocity between the recoiling BH and its host galaxy along the line of sight ($\approx 2650$ km\,s$^{-1}$ in the case of J0927+2943).

Alternative interpretations have been proposed for this object. \cite{Bogdanovic09} and \cite{Dotti09} have proposed two similar scenarios involving the presence of a sub-parsec BH binary. In this case the $z=0.697$ lines are associated with the broad line region of one component of the binary and with lower density gas in the binary sphere of influence, possibly associated with a gas stream fueling the active member (\citealt{Bogdanovic09}). In this framework, as for the recoiling case, the NELs at $z=0.7128$ set the reference frame of the host galaxy.

\cite{Heckman09} and \cite{Shields09} considered the less exotic interpretation of a chance superposition of two distinct galaxies within the 3$\arcsec$ angular size of the SDSS spectrograph fiber. In particular, \cite{Heckman09} suggested the presence of a single AGN at the redshift of the BELs that photoionizes the gas of a smaller galaxy falling toward it, responsible for the $z=0.7128$ NELs. \cite{Shields09} considered instead the case of two AGN, an unobscured type I AGN, responsible for the broad and narrow $z=0.697$ lines, and an obscured type II AGN associated with the other set of lines.

To discriminate among the different scenarios, we study the content and kinematics of the dense molecular gas, as traced by CO(2--1) emission. This allows us to check if the $z=0.7128$ lines actually trace the AGN host galaxy frame, as implied by the recoiling and binary scenario. We complement these observations with archival {\it HST} and dedicated ground--based optical spectroscopy and multi--band images, which allow us to further constrain the superposition model, and to investigate the properties of the quasar host galaxy and its environment.

Throughout the paper, we assume a standard $\Lambda$CDM cosmological model with $H_0$=70 \kms{} Mpc$^{-3}$, $\Omega_{\rm m}$=$0.3$, $\Omega_{\Lambda}$=$0.7$. At $z$=$0.7$, the corresponding angular scale implies that $1''\approx 7.15$ kpc. All magnitudes reported in this paper are in the AB photometric system.

\section{Observations}

\subsection{PdBI 2\,mm observations}

We observed the CO(2--1) transition in J0927+2943 using the IRAM Plateau de Bure Interferometer (PdBI). Observations were carried out in June 2012 (program W039) and in October 2013 (program X031) in the compact configuration (5Dq and 6Dq, respectively). The WideX band 2 (2mm) receiver was tuned to 134.5814 GHz in order to encompass the redshifted CO(2--1) frequency in all the redshift systems reported in the optical spectrum of J0927+2943. The PdBI primary beam size at this frequency is $37.4''$ wide, or 267\,kpc at the redshift of the system. The observations were centered on the optical coordinates of the QSO (R.A.: 09:27:12.66, Dec.: +29:43:44.2). Data were processed using the most recent release of the \textsf{GILDAS} suite. Data reduction was performed with \textsf{CLIC}. The bright radio sources 3C83, 0059+581, 0234+285, MWC349, 0923+392, 0851+202 were used as absolute flux calibrators, while 0851+202 and J0923+282 were observed as phase and amplitude calibrator. The final data cube consists of 9366 visibilities (7.8\,h on source, 6-antennas equivalent). Maps were created with the \textsf{MAPPING} software, adopting natural weighting of the visibilities. The resulting beam size is $4.55''\times3.86''$ ($33$\,kpc $\times$ $28$\,kpc). We reached a 1-$\sigma$ sensitivity of $0.43$ mJy\,beam$^{-1}$ per 100 \kms{} channel. Assuming a line width of 200 \kms{}, this sensitivity implies a 3-$\sigma$ detection limit on the CO(2--1) luminosity of $1.2\times10^9$ K\,\kms{}pc$^2$, or a 3-$\sigma$ limit on the molecular gas mass, $M_{\rm H2}$, of $\sim 10^9$ \Msun{} (assuming thermalized CO emission up to the $J$=2 transition and $\alpha_{\rm CO}=0.8$ \Msun{}[K\,\kms{}pc$^2$]$^{-1}$, as typically observed in high--$z$ QSOs, \citealt{Carilli13}). The continuum map, obtained by integrating line--free channels over 3800 \kms{}, has a 1-$\sigma$ RMS of 0.08 mJy\,beam$^{-1}$. Assuming that the dust spectral energy distribution is well described by a modified black body with dust emissivity index $\beta=1.6$ and temperature $T_{\rm dust}=30$ K \citep{Beelen06}, this implies a 3-$\sigma$ 
 limit on the IR luminosity of $1.3\times 10^{12}$ \Lsun{}, i.e., no continuum detection is expected, unless one of the sources had an unusually high (a few hundreds \Msun{}yr$^{-1}$) star formation rate.

\subsection{Calar Alto observations}

We spectroscopically observed J0927+2943 at optical wavelengths using the Cassegrain Twin Spectrograph (TWIN) on the Calar Alto 3.5m telescope on March 29, 2012. We discussed details of these observations in \citet{Decarli13}. Briefly, we adopted the T05 and T07 grisms on the blue and red arms respectively, providing spectral coverage in the ranges 3700--5000 \AA{} and 5500--11000 \AA{}, with a spectral resolution $\lambda / \Delta \lambda \approx 1300$ ($1''$ slit). The TWIN pixel scale in the slit direction is $0.56''$\,pxl$^{-1}$. The seeing during the observations, as measured from the FWHM of the QSO continuum along the slit, was $1.7''$. Data were reduced using standard \textsf{IRAF} routines. 

We also used the simultaneous multi--band capabilities of the BUSCA camera on the 2.2m telescope in Calar Alto in order to obtain deep optical images of J0927+2943. These observations were part of a program aimed at studying the galaxy environment of massive BH binary candidates, and will be presented in Mazzucchelli et al.~(in prep.). In brief, our broad--band observations span the whole optical window from {\it U} to {\it I} over a $12'\times12'$ field of view. The throughput curves of the dichroic mirrors, convolved with the detector efficiency, roughly correspond to those of the SDSS {\it u}, {\it r}, {\it i+z} and Johnson's {\it B} filters. We collected 5$\times$12 min frames, jithered by a few arcsec in order to effectively remove cosmic rays and bad pixels. Images were processed with our own pipeline \textsf{redbusca}, which is based on a collection of standard \textsf{IRAF} procedures. The astrometric solution of the final images was obtained with \textsf{astrometry.net} \citep{Lang10}. We derived the photometric Zero Points by comparing the photometry of stars in the field with the SDSS catalogs, after applying the required corrections to account for the slightly different throughput curves (see Mazzucchelli et al.~for details). The 5-$\sigma$ detection limits for a point source are $25.38$, $25.01$, $25.13$, $24.21$ mag in {\it U}, {\it B}, {\it R}, and {\it I} respectively. The seeing in {\it R} band was $1.55''$. 

\subsection{{\it HST} observations}

We complemented our photometric observations with {\it HST} archival data (program ID: 11624). J0927+2943 was observed with the Advanced Camera for Surveys (ACS) + filter F606W and with the Wide Field Camera 3 (WFC3) in the IR filter F110W in May 2010. The time on source was 1000 s and 1997 s respectively. At $z\approx 0.7$, the two filters roughly correspond to rest frame {\it u} and {\it r+i}. We refer to the science products obtained with the official {\it HST} \textsf{MultiDrizzle} pipeline. 

The ACS observations were collected with the Wide Field Camera, which has a pixel scale of $0.05''$\,pxl$^{-1}$ and a field of view of $202''\times 202''$. The WFC3 IR observations have a pixel scale of $0.13''$\,pxl$^{-1}$ and a field of view of $136''\times 123''$. We estimate that the 3-$\sigma$ surface magnitude limits in the two images are $26.1$ mag\,arcsec$^{-2}$ and $26.4$ mag\,arcsec$^{-2}$ in the F606W and F110W bands, respectively.

\section{Results}

\begin{figure}
\includegraphics[width=0.99\columnwidth]{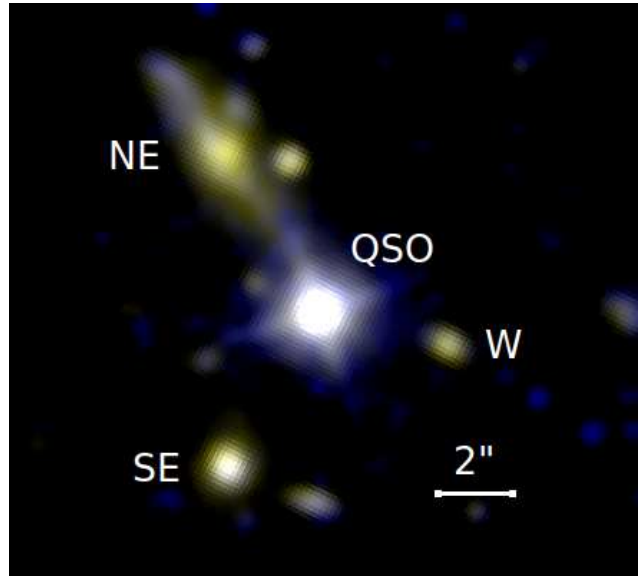}
 \caption{{\it HST} F606W (blue) and F110W (yellow) images of J0927+2943. North is up, East is on the left--hand side. The ACS F606W image was degraded to match the resolution and sampling of the WFC3 IR F110W image. The QSO and the main surrounding galaxies are labeled. The host galaxy of the QSO is unresolved at {\it HST} resolution. The NE companion shows a disturbed morphology, elongated in the direction of the QSO, suggesting that an interaction may be ongoing.}\label{fig_hst}
\end{figure}

In the following, we summarize our results for the QSO J0927+2943, for the closest nearby companion galaxies (as dubbed in Fig.~\ref{fig_hst}), and for the large scale environment.

\begin{figure*}
\includegraphics[angle=-90, width=0.99\textwidth]{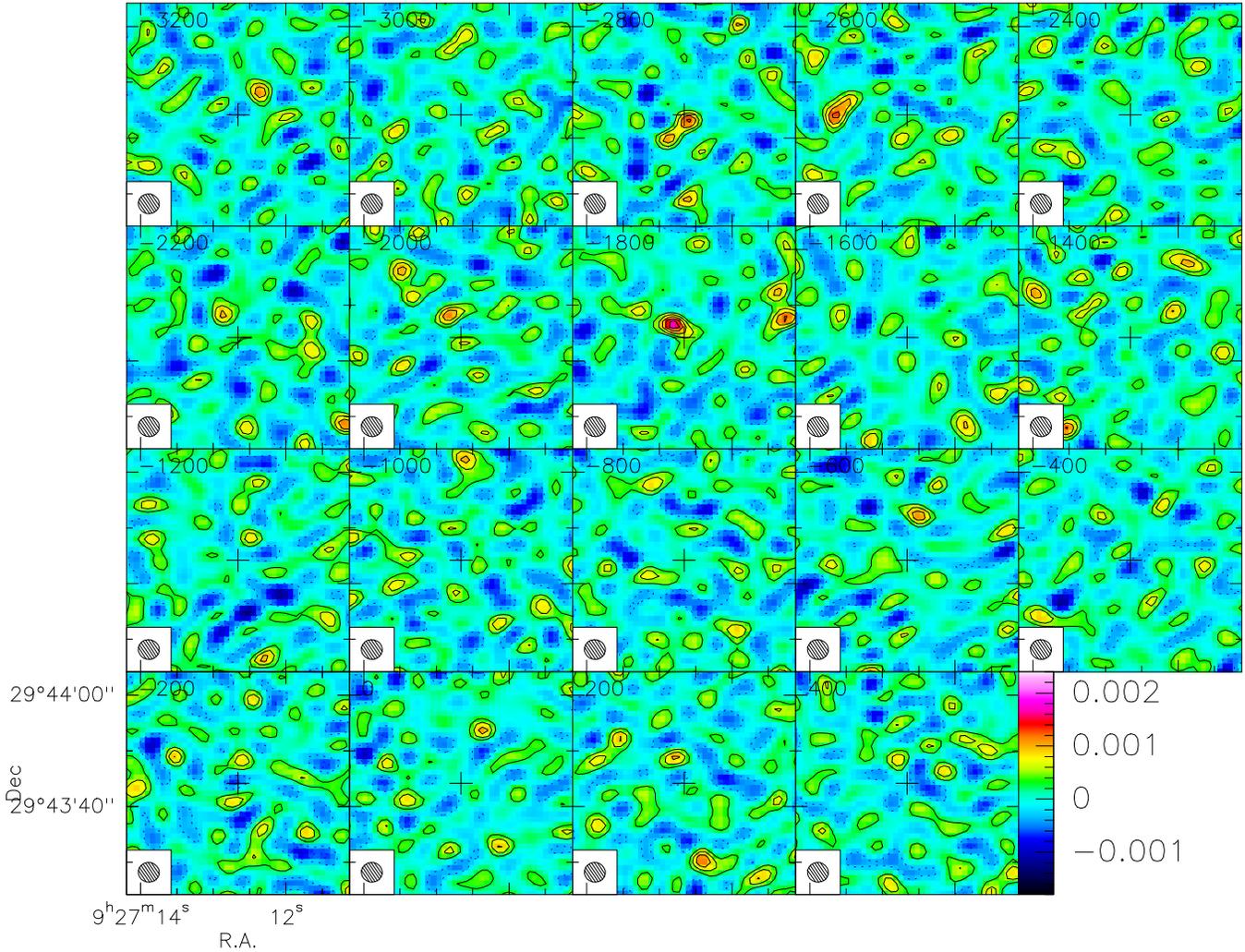}
 \caption{Channel maps of our 2mm CO(2--1) observations of J0927+2943. The data cube has been rebinned into 200\,\kms{} wide channels. The color scheme map the flux density in units of Jy\,beam$^{-1}$. Black solid (blue dotted) contours are +1, +2, +3, \ldots (-1, -2, -3, \ldots) $\sigma$ spaced. The beam of our observations is shown as a shaded ellipse in the bottom-left corner of each panel. The central cross marks the pointing centre of our observations, i.e., the QSO position. Clear line detections are apparent in the -2800\,\kms{} channel and in the range -2200\,\kms{} -- -1800\,\kms{}, and are associated with the QSO host galaxy and SE companion, and with the NE companion galaxy respectively. No line is reported at $v\approx 0$, consistent with the redshift of the red NEL system.}\label{fig_chnmap}
\end{figure*}

\subsection{Molecular emission in J0927+2943}\label{sec:2mm}

\begin{figure}
\includegraphics[width=0.99\columnwidth]{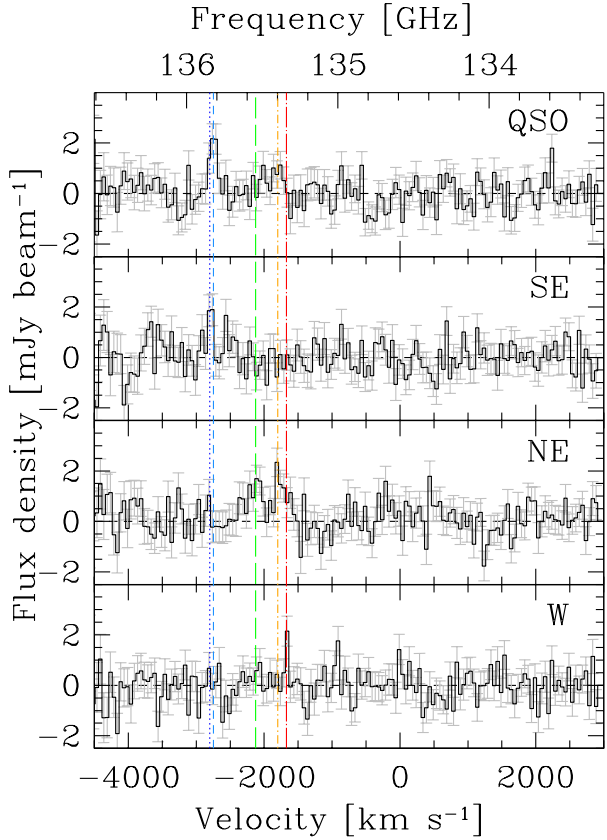}
 \caption{Spectra of the 2mm emission of the QSO and the three companion galaxies within 50 kpc in projected distance. The spectra have been rebinned into 50 \kms{} wide channels. The velocity zero point is set by the CO(2--1) frequency at the redshift of the bright set of narrow emission lines in the optical spectrum (`red NEL', see Fig.~\ref{fig_caha_spc}). No emission is detected at this redshift. On the other hand, we report the clear detection of CO(2--1) at the redshift of the blue set of narrow lines + broad lines ($z$=$0.69744$, blue short-dashed line). A fainter emission is also reported at the redshift of the intermediate system associated with the NE companion (orange dot-dashed line). }\label{fig_spc}
\end{figure}

We use our PdBI 2mm CO observations to firmly pin down the redshifts of the QSO host galaxy and of the closest companion galaxies. In Fig.~\ref{fig_chnmap} we show the channel map of our PdBI 2mm observations, in bins of 200\,\kms{}. The spectra extracted at the position of the QSO, and of the closest companion galaxies (all within 50\,kpc, i.e., well within the primary beam of our observations) are shown in Fig.~\ref{fig_spc}. 

We find a clear line spatially consistent with the position of the QSO. The line peaks at $135.82\pm0.06$ GHz. The associated CO(2--1) redshift is $0.69744\pm0.00008$, consistent with the redshift of the broad lines and the `blue' system of narrow lines in the optical QSO spectrum. The line flux is $0.32\pm0.05$ Jy\,\kms{}, yielding a line luminosity $L_{\rm c} = (8.1\pm1.3) \times10^5$ \Lsun{} or $L'=(2.1\pm0.3)\times 10^9$ K\,\kms{}pc$^2$. Assuming that CO emission is thermalized, and a ULIRG-like CO--to--H$_2$ conversion factor ($\alpha_{\rm CO}=0.8$ \Msun{} [K\,\kms{}pc$^2$]$^{-1}$), we derive a molecular gas mass of $\sim1.7\times10^9$ \Msun{}. Such an estimate could be up to a factor $\sim 8$ higher in the case of low CO excitation and Milky Way--like values of $\alpha_{\rm CO}$. Such molecular gas reservoir is commonly observed in star forming galaxies and QSOs \citep{Carilli13}, and points towards a galactic--scale gas reservoir. This unambiguously sets the redshift of the QSO host galaxy.

No CO line is found at the redshift of the red NEL system. For a line width of 200 \kms{}, roughly consistent with the line width of the nebular lines, we derive a 3-$\sigma$ limit of $0.18$ Jy\,\kms{}, or $L'=1.2\times10^9$ K\,\kms{}pc$^2$. This implies that the molecular gas reservoir associated with this redshift system is not particularly large ($M_{\rm H2}\lsim 10^9$ \Msun).

\begin{figure}
\includegraphics[width=0.99\columnwidth]{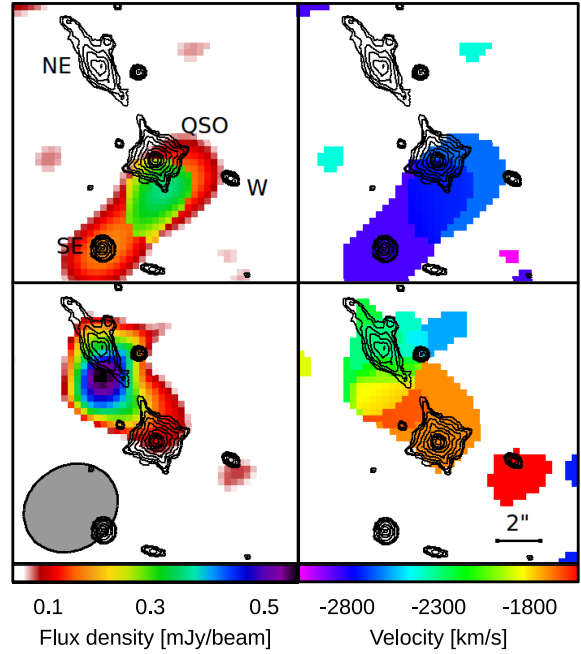}
 \caption{{\em Top row ---} Zeroth ({\em left}) and first ({\em right}) moment maps (color scale) of the QSO and of the SE companion in J0927+2943. The black contours are logarithmically--spaced isophotes of the {\it HST}/WFC3 F110W image (see Fig.~\ref{fig_hst}). The main components of the system are labeled. The moment maps are integrated between -2900 \kms{} and -2600 \kms{}. The QSO is clearly detected, while the detection of the companion source is only marginal. {\em Bottom row ---} Same for the NE and W companions. The ellipse in the bottom--left corner shows the resolution element of our PdBI 2mm observations. The NE galaxy is clearly detected. Its CO emission is spatially resolved towards the QSO and shows a velocity gradient ($\sim600$ \kms{} peak to peak along the NE--SW axis). A tentative, narrow CO line is reported at the position of the W companion.}\label{fig_mom}
\end{figure}

Our PdBI observations also reveal an extended  CO(2--1) emission associated with the NE companion galaxy, elongated towards the QSO and even overlapping with it. The CO line profile is very broad ($\sim 1000$ \kms{}; see Fig.~\ref{fig_spc}). It is unclear whether the large width of the CO line in the NE is due to a single object or to two distinct galaxies. The line peaks at 135.5 GHz ($z=0.701$), and its integrated flux is $\sim 1.0$ Jy\,\kms{} (the exact value depending on the region considered for the flux measurement). The associated line luminosity is $L_{\rm c}\approx2.6\times10^6$ \Lsun{} or $L'\approx6.6\times10^9$ K\,\kms{}pc$^2$. Assuming the same conversion factors adopted for the QSO host galaxy, we infer a molecular gas mass of $M_{\rm H2}\approx5\times10^9$ \Msun{}. A clear velocity gradient is observed in the CO(2--1) first moment map (see Fig.~\ref{fig_mom}), in the sense that the part of the galaxy that is more distant from the QSO is blueshifted and the side leaning towards the QSO is redshifted. The velocity map, derived by integrating between -2500 \kms{} and -1600 \kms{} (in the rest frame of the red set of NELs in Fig.~\ref{fig_caha_spc}; see also Fig.~\ref{fig_spc}), suggests that the peak--to--peak gradient is $\sim 600$ \kms{}. A hint for such a gradient velocity gradient is also reported in the 2-D optical spectrum of J0927+2943 (see Fig.~\ref{fig_caha_spc}), although the significance is modest. 

We also report a tentative (3-$\sigma$) CO(2--1) line detection in the SE companion galaxy (see Fig.~\ref{fig_chnmap}--\ref{fig_mom}). The line is detected over 4 independent channels (at 25 \kms{} wide binning). The CO line is centered at the position of the optical counterpart (see Fig.~\ref{fig_mom}). It peaks at $135.84\pm0.01$ GHz ($z=0.6973\pm0.0001$). Its flux is $0.20\pm0.6$ Jy\,\kms{}, implying a CO(2--1) luminosity of $L_{\rm c}=(4.8\pm1.4)\times10^5$ \Lsun{} or $L'=(1.2\pm0.4)\times10^9$ K\,\kms{}pc$^2$. The line peaks at -2805 \kms{} with respect to the red NEL system in the QSO spectrum, i.e., 50 \kms{} bluewards of the QSO host galaxy.


\subsection{$L_*$ and $M_*$ of the QSO host galaxy}\label{sec:host}

\begin{figure}
\includegraphics[width=0.99\columnwidth]{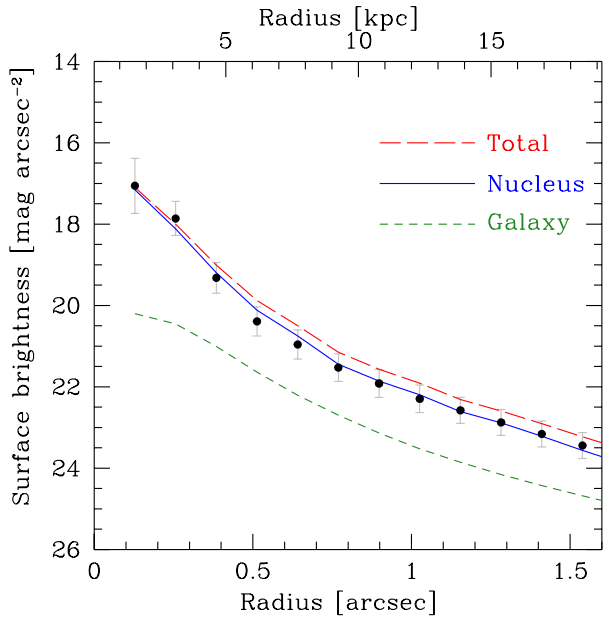}
 \caption{Radial light profile of the QSO in J0927+2943, as measured on the {\it HST}/WFC3 F110W image shown in Fig.~\ref{fig_hst}. The light profile is perfectly consistent with a pure Point Spread Function (solid line). Our simulations suggest that the brightest galaxy model which is not ruled out by the observations has a total F110W flux of $>20.0$ mag, a de Vaucouleurs profile, and an effective radius of $\lsim1$ kpc (short-dashed line). The total emission of the nucleus + the brightest allowed galaxy is shown as a long-dashed line.}\label{fig_rad_prof}
\end{figure}

We analyse the available {\it HST} images of the field in order to put constraints on the stellar emission from the QSO host galaxy. Ground--based images lack of the required angular resolution for this study. The {\it HST}/ACS F606W image, on the other hand, has the best available angular resolution, but it maps the rest--frame $U$-band emission, where the contrast between the nuclear emission and the stellar continuum is less favourable. Therefore, we base our analysis on the {\it HST}/WFC3 IR F110W image. The observed light profile is examined out to a radius of $1.6''$ (at larger radii, the contamination from the NE companion and from another, independent blob on the Eastern side hinder our analysis). The light profile is fully consistent with a Point Spread Function, the model of which we derived from the images of field stars (see Fig.~\ref{fig_rad_prof}). 

We simulated various galaxy models assuming de Vaucouleurs or exponential disc profiles, effective radii ranging between 1 and 5 kpc, and various fluxes in order to evaluate the limit luminosity of the host galaxy stellar emission \citep[following the approach presented in][]{Decarli12}. We found that the brightest allowed galaxy has a de Vaucouleurs profile, an effective radius of $\lsim1$ kpc and total F110W magnitude $>20.0$ mag. Assuming the Elliptical galaxy template by \citet{Mannucci01} for $k$-correction, we derive a limit on the rest-frame $R$-band luminosity of $M_R>-22.3$ mag. In the hypothesis of a relatively old stellar population, this limit implies a stellar mass of $\lsim 2.5\times 10^{11}$ \Msun{} (younger stellar populations would imply a stricter limit on the stellar mass). This is the largest stellar mass allowed by the available data (i.e., our limit is defined in a very conservative way). Such limit is not particularly stringent: For instance, assuming the black hole mass estimate by \citet{Komossa08} ($M_{\rm BH}\approx 6 \times 10^8$ \Msun), we infer a $M_{\rm BH}/M_{\rm host}$ ratio of $>0.002$, still in agreement with the BH--host galaxy relations observed at low $z$. 

From the analysis of the {\it HST} images, the NE companion appears as an edge--on disc galaxy, aligned with and elongated towards the QSO along the NE--SW direction. Its morphology appears distorted in the {\it HST} images (see Fig.~\ref{fig_hst}). 


\begin{figure}
\includegraphics[width=0.99\columnwidth]{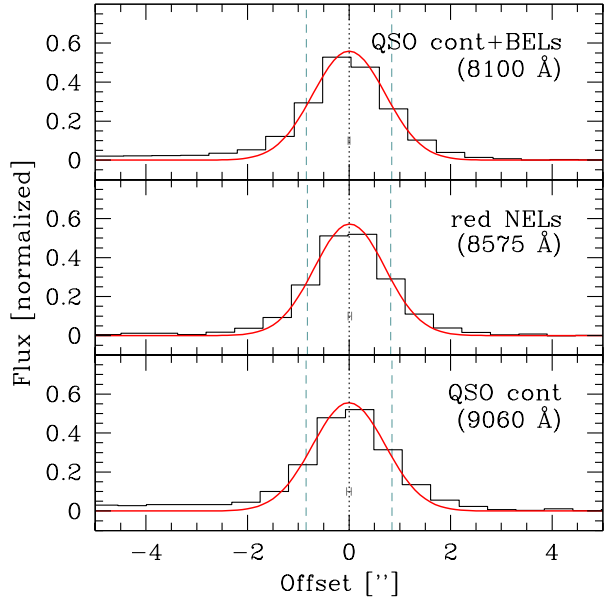}
 \caption{Light profile of J0927+2943 extracted along the slit from our TWIN observations (see Fig.~\ref{fig_caha_spc}). The light profile is extracted at the wavelengths corresponding to the QSO continuum plus H$\beta$ broad emission line; to the \Oiii{} 5008 \AA{} in the red-NEL system; and to the QSO continuum in the $i$ band. The line profile is fitted with a gaussian curve. The offset is defined with respect to the QSO continuum position; negative (positive) offsets are NE (SW) of the QSO. The error bars mark the 3-$\sigma$ confidence range on the peak position of the profile. Dashed, vertical lines mark the FWHM of the light profile. The red-NEL system is fully consistent with the peak position of the QSO continuum, and has the same width (= unresolved at the seeing resolution).}\label{fig_align}
\end{figure}

\subsection{Alignment between the red-NEL system and the QSO}\label{sec:align}

The TWIN spectrum shows good alignment along the slit between the red-NEL system and the QSO continuum (see Fig.~\ref{fig_caha_spc}). In Fig.~\ref{fig_align}, we compute the position of the QSO continuum along the slit blue- and redwards of the \Oiii{} lines in the red-NEL system. We linearly interpolate in order to estimate the QSO position at the wavelengths of the red-NEL \Oiii{}. We compare it with the position along the slit of the red NELs. The brightness of the red-NEL \Oiii{} lines is such that we can easily achieve very high centering accuracy compared with the seeing conditions. The \Oiii{} of the red-NEL system peaks $0''.014\pm0''.012$ $\approx (100\pm 90)$\,pc (1-$\sigma$), i.e., no significant offset is reported.

The FWHM of the \Oiii{} 5008\AA{} line in the red-NEL system is $1.67''$, fully consistent with the one of the QSO continuum. This disagrees with previous claims of a spatially--extended \Oiii{} emission along the NE--SW direction \citep{Vivek09}.

%

\begin{figure}
\includegraphics[width=0.99\columnwidth]{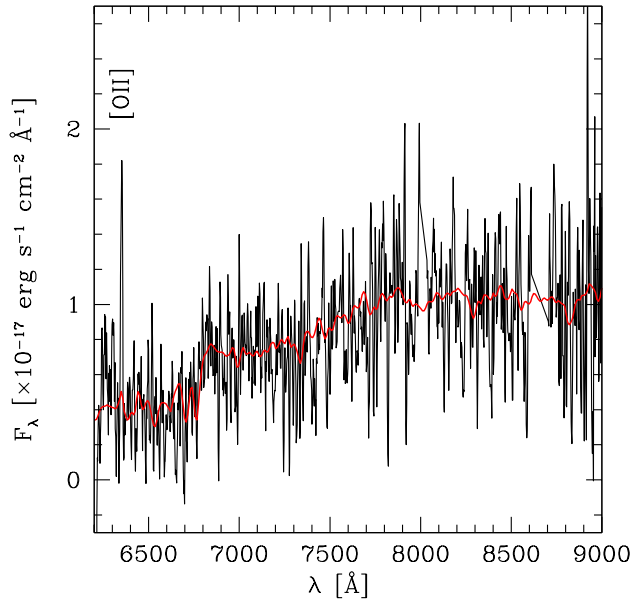}
 \caption{Optical spectrum of the SW companion (see also Fig.~\ref{fig_caha_spc}). The observed spectrum clearly shows the Ca break and the \Oii{} line. The thick, red line is the S0/Sa composite template derived from \citet{Kinney96} that was used for the redshift estimate. The continuum redshift is fully consistent with the one infered from the \Oii{} line.}\label{fig_SWspec}
\end{figure}

\subsection{SED of the close companions}\label{sec:mstar}

We use the galaxy SED fitting tool \textsf{MAGPHYS} \citep{Charlot00,BC03,EdC08} to model the available photometry (BUSCA+{\em HST}) of the galaxies in a close projected distance from the QSO. The NE companion has a stellar mass $M_*=(7.1_{-1.5}^{+2.2})\times 10^{10}$ \Msun{} and a star formation rate SFR=$30\pm20$ \Msun{}yr$^{-1}$. The rest-frame optical/UV emission of this source show significant extinction. Assuming energy balance between UV absorption and re-emission at IR wavelengths, we derive an extrapolated IR luminosity of $\sim 10^{11}$ \Lsun{}, i.e., in the LIRG regime (and yet consistent with a non-detection in our 2mm continuum map). Within the uncertainties in the conversion factors, the $M_{\rm H2}/M_*\approx 0.07$ ratio is in agreement with what typically observed in `main sequence' galaxies at this redshift \citep[see, e.g., Fig.~9 in][]{Carilli13}. For the SE companion, our \textsf{MAGPHYS} fit of the available photometry suggests that the galaxy has a stellar mass $M_*=(7_{-3}^{+4})\times 10^9$ \Msun{} and a SFR=$8_{-2}^{+8}$ \Msun{}yr$^{-1}$. Finally, we estimate a stellar mass $M_*=(1.4\pm0.4)\times 10^{11}$ \Msun{} and a star formation rate SFR=$15_{-8}^{+15}$ \Msun{}yr$^{-1}$ for the SW companion.

\subsection{Large-scale environment}\label{sec:env}

Our BUSCA images of J0927+2943 cover $\approx 26.86$ Mpc$^2$ (in projected distance) around the QSO, i.e., they are sufficiently large to encompass the extent of any plausible galaxy cluster at the QSO redshift. Our $R$-band image is the deepest large-scale image available of the field (a factor $\sim 14$ deeper than SDSS $r$ band). We identify sources in this image using \textsf{SExtractor} \citep{Bertin96}. Out of them, the majority (1254) are galaxies (based on the comparison between PSF and Sersic magnitudes). For each source, we computed photometric redshifts following the Bayesian approach described in Mazzucchelli et al.~(in prep.). Namely, we compare the color information of each source (including its uncertainties) with a library of galaxy templates at various redshifts. The grid of templates is obtained from a linear interpolation of various empirical galactic templates from \citet{Kinney96}, with various degrees of intrinsic reddening ($E$($B$-$V$)=0--0.2 mag) based on the extinction law by \citet{Calzetti94}. Our priors include information on the luminosity function of galaxies \citep[adapted from][]{Gavazzi10} and on the volume of universe probed by our images as a function of $z$. The method computes probability density functions (PDFs) of each source being described by a certain galaxy template at a given redshift. We take the median value of the PDFs marginalized over the galaxy template as our best estimate of $z_{\rm phot}$, and the 14\% and 86\% interquartiles of the PDF as the 1-$\sigma$ confidence levels. The typical uncertainties in the photometric redshifts derived in this way are $\delta z \approx 0.09$.

A hint for the robustness of our photometric redshifts comes from the TWIN spectroscopy and the PdBI 2mm CO(2--1) observations of J0927+2943: The TWIN slit was oriented so that we encompassed both the NE and the SW companions. The former has an optical spectroscopic redshift $z=0.701$ (see Fig.~\ref{fig_caha_spc}), fully consistent with the one derived from the CO(2--1) observations (see Fig.~\ref{fig_spc}). Our photometric redshift estimate is $z_{\rm phot} = 0.78_{-0.04}+^{+0.05}$, i.e., in agreement within $1.9$ $\sigma$. The photometric redshift of the SE companion is significantly less secure, $z_{\rm phot}=1.1_{-0.5}^{+0.2}$, yet consistent with our tentative CO detection ($z=0.6973\pm0.0001$). Finally, the SW companion is a massive galaxy found $\sim21.4''\approx 153$ kpc South-West of the QSO. Our analysis of the BUSCA images suggests the galaxy has a photometric redshift $z_{\rm phot}=0.76_{-0.06}^{+0.05}$. The optical spectrum of this galaxy is shown in Fig.~\ref{fig_SWspec}. The shape of the continuum emission suggests that the galaxy has a relatively red stellar population. The Calcium break feature pins down the continuum redshift at $0.7043 \pm0.0009$. The only emission line associated with this galaxy is at $6351.0\pm0.5$ \AA{}, and is identified with the \Oii{} line at a redshift $z_{\rm [OII]}=0.7039 \pm 0.0005$, consistent with both the redshift inferred from the continuum spectrum and with our photometric redshift estimate. 

\begin{figure*}
\includegraphics[width=0.49\textwidth]{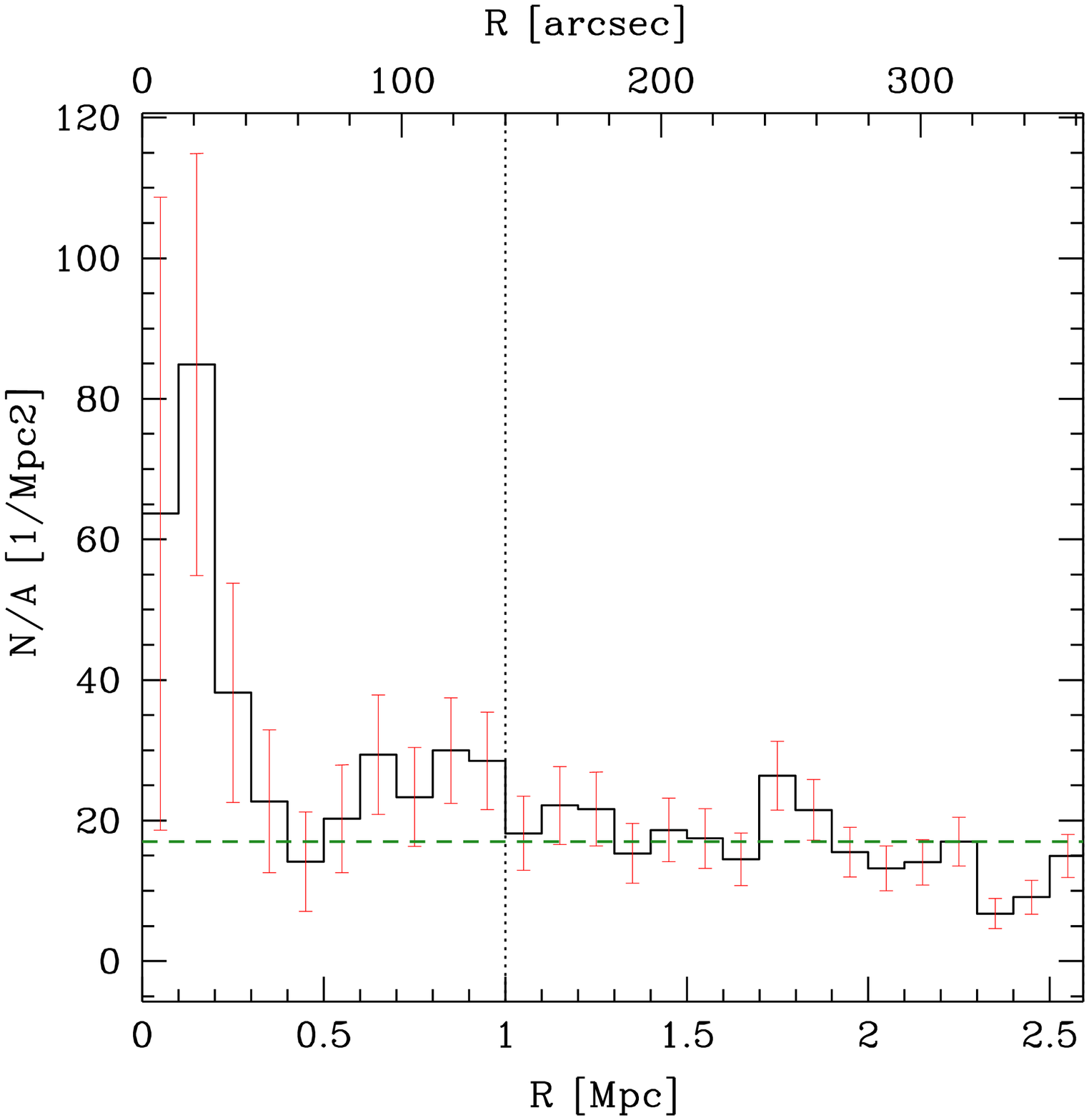}
\includegraphics[width=0.49\textwidth]{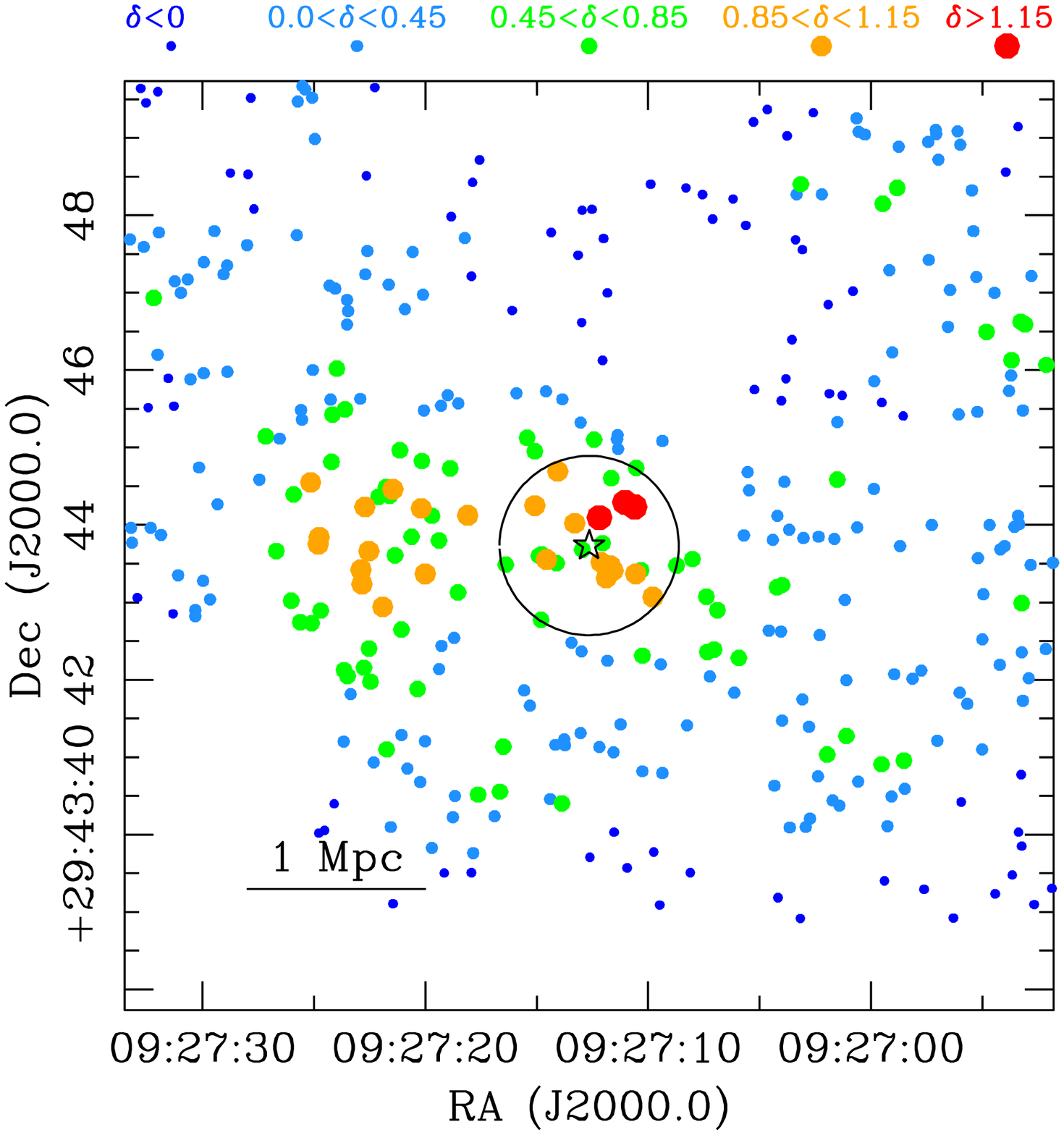}\\
 \caption{{\em Left --} Surface density of galaxies as a function of the radial distance from the QSO in J0927+2943, based on our $R$-band selected sample, trimmed to consider only the sources with photometric redshift consistent with the QSO. An excess over the median background (computed at 1\,Mpc $< R <$ 2.5\,Mpc and highlighted with a green horizontal line) is apparent at separations $<0.5$ Mpc. Error bars show the Poissonian uncertainties in each bin. {\em Right --} Map of the field around J0927+2943. Only the central $4.4\times4.4$ Mpc$^2$ are shown. The QSO position is marked with a star in the centre. Only galaxies with photometric redshift consistent with the one of the QSO are shown, color coded according to the local galactic density $\delta$ (see text for details). A group of galaxies associated with the QSO is apparent in the centre of the frame. For reference, the circle marks a distance of 0.5 Mpc from the QSO.}\label{fig_envir}
\end{figure*}

In order to investigate the presence of a galaxy overdensity around the QSO, we restrict our analysis to the sample of sources with photometric redshifts consistent (within 1 $\sigma$) with the one of the QSO host galaxy. In Fig.~\ref{fig_envir}, {\it left}, we plot the surface density of such galaxies as a function of projected distance from the QSO. An excess is reported at radii $<0.5$ Mpc. We estimate the level of background by computing the median value of the surface density at radii between 1 and 2.5 Mpc in projected distance from the QSO. This background is likely associated with large-scale structures at redshifts similar to the one of the QSO, or fore/background sources with highly uncertain photometric redshifts. This allows us to estimate that the excess observed at small radii ($<$1 Mpc) accounts for $35\pm10$ galaxies. Further support to the presence of a small galactic overdensity is shown in Fig.~\ref{fig_envir}, {\em right}. Here we defined a ``local density'' parameter as follows: For each  source, we count the number of companion galaxies within $<0.5$ Mpc and with $z_{\rm phot}$ consistent within 1 $\sigma$ with the one of the QSO. This number is divided by the area corresponding to a radius of 0.5 Mpc, thus obtaining the local surface density of galaxies $\Sigma_{\rm gal}$. The local density parameter is then obtained as: $\delta=[\Sigma_{\rm gal} - \langle \Sigma_{\rm gal} \rangle]/\langle \Sigma_{\rm gal} \rangle$., where $\langle \Sigma_{\rm gal} \rangle$ is the median background surface density ($15\pm1$ galaxies per Mpc$^2$). A clump  is apparent in the center of our image. These findings support the evidence of a galaxy overdensity, as indicated also by the presence of a second QSO 125\,kpc away (in projected distance) from J0927+2943 \citep{Decarli10}.

In order to gauge the overdensity, we compare these numbers with the number of galaxies that we expect to detect if we redshifted the Coma cluster to $z=0.7$, following our previous analysis \citep{Decarli09}. We take the luminosity function of Coma from \citet{Gavazzi10}, considering only their ``Ultra High Density'' regime, and scale it down to match an area with projected radius = 1 Mpc (as adopted here for J0927+2943). After taking into account filter- and $k$-corrections, we integrate the Coma luminosity function down to the corresponding limit magnitude in our observations. We thus estimate that, if a galaxy cluster comparable with Coma was present, we would detect $\sim 130$ galaxies. The excess of galaxies reported here is thus $0.28\pm0.08$ times the one expected for a Coma-like cluster. These estimates are necessarily conservative (as our typical uncertainties on $z_{\rm phot}$ are large). They confirm our previous estimates based on $K_{\rm s}$-band selected galaxies in the field \citep{Decarli09}, and hint toward the presence of a galaxy group rather than a massive cluster.

\section{Discussion}

The observational data presented in this work put strong constraints on the nature of J0927+2943. Here we discuss whether the three models proposed to date for this object are consistent with these new observations.

We detected a clear CO(2--1) emission line spatially consistent with the position of the QSO. The redshift of the molecular line, tracing the host galaxy rest frame, is consistent with that of the blue-shifted broad and narrow lines (i.e., the blue system) observed in the optical spectrum. No molecular line emission is detected at the redshift of the second, red narrow line system. As described in the Introduction, both the recoiling AGN and the binary model assume the blue system to be off-set with respect to the host galaxy redshift which is considered to coincide with that of the red system. Hence the PdBI 2mm observations discussed in Section~\ref{sec:2mm} definitely disprove these two models.

The other possible scenario proposed for J0927+2943 is that of a superposition. In this case the blue system is associated with the QSO while the red system originates in a second distinct structure (e.g. a second AGN or a smaller galaxy illuminated by the QSO) along the same line of sight. In order to boost the otherwise negligible probability of such a chance superposition, the QSO is assumed to inhabit a rich galaxy cluster (\citealt{Heckman09}, \citealt{Shields09}). Under this assumption \cite{Shields09} reported the probability of a chance superposition of two AGN within $\lsim 1''$ (in order to be observed in the same SDSS fiber) of $\sim 10^{-4.3}$, so that few chance superpositions are to be expected in the SDSS quasar catalogue. Such a prediction has been confirmed through an analysis of the results of the Millennium Run (\citealt{Springel05}) by \cite{Dotti10}. However these authors find that no high relative velocity chance superposition is expected at $z\gsim 0.4$, disfavouring the superposition scenario in the specific case of J0927+2943.  The drop in the superposition probability at $z\sim 0.4$ is related to the low abundance of galaxy clusters at high redshift, as predicted by the hierarchical model of structure formation.

The TWIN spectrum presented here further constrains the superposition scenario. The red and blue systems are found to be superimposed within $\approx 0.012''$. Hence, the probability estimated by \cite{Shields09} has to be rescaled by a factor $\sim 10^{-4}$. Within such a small area $\ll 1$ superimposed AGN pair is expected in the SDSS database, even considering the effect of clusters. In the {\it HST} image a close (in projection) companion disc galaxy (the NE companion in Fig.~\ref{fig_hst}) is clearly visible. We note that the presence of a companion galaxy with a disturbed morphology in high resolution optical images has been predicted in \cite{Heckman09}. If involved in a close gravitational interaction, the disc galaxy would be elongated toward the QSO thus boosting the chance superposition probability. However, from the optical and 2mm spectra we derive that the companion galaxy is associated with a third faint set of emission lines at an intermediate redshift, originally reported by \cite{Shields09}. The redshift difference between this third line system and the blue one (i.e., the QSO) implies a relative velocity of only $\sim 850$ \kms along the line of sight, hence the companion galaxy cannot be responsible for the red line system given that the observed velocity difference is instead $\approx 2650$ \kms. 

An alternative option is that the red line system is associated with material stripped from the companion and falling toward the QSO. We can use simple energy conservation arguments to show that this is unlikely, by estimating the maximum velocity such a cloud would acquire in the fall. Generously we optimize the conditions (i.e., maximizing the final velocity of the cloud) by assuming that the cloud falls straight toward the QSO subject only to the gravitational potential of the QSO host (e.g. neglecting the deceleration due to the potential well of the companion) and that the initial separation between the two galaxies is $\gg$ than the QSO host scale radius. Under these assumptions we can approximate the energy conservation law as follows: 
\begin{equation}
\frac{v_{\rm fin}^2}{2} \sim G\frac{M_{\rm QSO,h}}{a_{\rm QSO,h}}+\frac{v_{\rm in}^2}{2}, 
\end{equation}
where $M_{\rm QSO,h}$ and $a_{\rm QSO,h}$ are the mass and scale radius of the QSO host galaxy, $v_{\rm in}$ is the initial velocity of the cloud when it is tidally stripped from the companion galaxy, and $v_{\rm fin}$ is the cloud velocity after having been accelerated by the QSO host.  To further maximize $v_{\rm fin}$ we set $v_{\rm in}=1075$ \kms, i.e. the maximum relative velocity between the QSO host and the companion observed in the sub-mm spectrum. Despite these extreme assumptions, the QSO host should still have an unrealistically high mass of $M_{\rm QSO,h} \approx 6.8 \times 10^{10}$ \Msun{} within $a_{\rm QSO,h}\approx 100$ pc or equivalently $M_{\rm QSO,h} \approx 6.8 \; 10^{11}$ \Msun within $a_{\rm QSO,h}\approx 1$ kpc in order to accelerate the cloud to the observed velocity difference between the red and blue systems in the optical spectra\footnote{We checked this result with a more detailed model, assuming the QSO host to follow an Hernquist profile and determining its mass and scale radius from the \Mbh-host galaxy scaling relations (\citealt{Haring04}, \citealt{Gultekin09}). The maximum velocity achievable in this case is only $\approx 1500$ \kms, about 1000 \kms less that the observed velocity difference in the optical spectra.}.

Finally, in Section \ref{sec:env} we performed a clustering analysis of the QSO field, detecting a galaxy overdensity consistent with a rich galaxy group rather than a galaxy cluster. As discussed above, theoretical arguments as well as the new observations presented here seems to disfavour the superposition scenario. Still, this is the only proposed scenario that has not yet been unambiguosly ruled out by a key observational test. The projected separation of a possible satellite superimposed on the QSO has to be less than $0.1''$ making it extremely difficult to resolve the presence of such a system in the near future. On the other hand, a spectroscopic follow-up of the galaxies with a photometric redshift consistent with that of the QSO would improve the characterization of the galaxy overdensity potential. This could probe the possible presence of a galaxy cluster with an anomalously high dark matter mass to galaxy number ratio needed to justify a high relative velocity (\citealt{Decarli09}). However if the velocity dispersion of the galaxies would turn out to be significantly lower than $\approx 2500$ \kms, consistent with the presence of a group, the superposition scenario would then be ruled out as well as the recoil and binary scenarios, leaving the nature of J0927+2943 still an open issue.


\section*{Acknowledgments}

We thank the anonymous referee for helpful comments. We thank F.~Walter for his precious comments on the manuscript. We thank C.~Feruglio, S.~K\"{o}nig, and R.~Neri for their valuable help with the observations and reduction of PdBI data, and the CAHA astronomers who collected our optical data. Support for RD was provided by the DFG priority program 1573 ``The physics of the interstellar medium.''. CM is supported by the EXTRA program at Universit\`{a} degli Studi di Milano--Bicocca. MV acknowledges funding support for this research from a Marie Curie FP7-Reintegration-Grants within the 7th European Community Framework Programme (PCIG10-GA-2011-303609).

This work is based on observations made with the IRAM Plateau de Bure Interferometer, at the Centro Astron\'{o}mico Hispano Alem\'{a}n (CAHA), and with the NASA/ESA Hubble Space Telescope. IRAM is supported by INSU/CNRS (France), MPG (Germany) and IGN (Spain). CAHA is operated jointly by the Max-Planck Institut f\"{u}r Astronomie and the Instituto de Astrofisica de Andalucia (CSIC). HST data were obtained from the Data Archive at the Space Telescope Science Institute, which is operated by the Association of Universities for Research in Astronomy, Inc., under NASA contract NAS 5-26555. These observations are associated with program \#11624.



\bsp

\label{lastpage}

\end{document}